\documentclass[preprint,journal]{vgtc}            

\onlineid{1346}

\vgtccategory{Research}

\title{Leveraging Historical Medical Records as a Proxy via Multimodal Modeling and Visualization to Enrich Medical Diagnostic Learning}
\author{%
\authororcid{Yang Ouyang}{0009-0000-5841-7659}, \authororcid{Yuchen Wu}{0009-0005-8333-4405}, \authororcid{He Wang}{0009-0003-2550-6139}, \authororcid{Chenyang Zhang}{0009-0003-1116-4895}, \authororcid{Furui Cheng}{0000-0003-2329-6126}, \\Chang Jiang, Lixia Jin, Yuanwu Cao, and \authororcid{Quan Li}{0000-0003-2249-0728}
}

\authorfooter{
  \item
  	Yang Ouyang, Yuchen Wu, He Wang, and Quan Li are with the School of Information Science and Technology, ShanghaiTech University, and Shanghai Engineering Research Center of Intelligent Vision and Imaging, China. Quan Li is the corresponding author. E-mail: \{ouyy,wuych3,wanghe1,liquan\}@shanghaitech.edu.cn.
  \item
  	Chenyang Zhang is with the Department of Computer Science, University of Illinois at Urbana-Champaign. E-mail: zhang414@illinois.edu.
  \item Furui Cheng is with the Department of Computer Science, ETH Zürich. E-mail: furui.cheng@inf.ethz.ch.
  \item Chang Jiang, Lixia Jin, and Yuanwu Cao are with Zhongshan Hospital Fudan University. E-mail: cjiang\_fdu@yeah.net, \{jin.lixia,cao.yuanwu\}@zs-hospital.sh.cn.
}

\abstract{Simulation-based Medical Education (SBME) has been developed as a cost-effective means of enhancing the diagnostic skills of novice physicians and interns, thereby mitigating the need for resource-intensive mentor-apprentice training. However, feedback provided in most SBME is often directed towards improving the operational proficiency of learners, rather than providing summative medical diagnoses that result from experience and time. Additionally, the multimodal nature of medical data during diagnosis poses significant challenges for interns and novice physicians, including the tendency to overlook or over-rely on data from certain modalities, and difficulties in comprehending potential associations between modalities. To address these challenges, we present \textit{DiagnosisAssistant}, a visual analytics system that leverages historical medical records as a proxy for multimodal modeling and visualization to enhance the learning experience of interns and novice physicians. The system employs elaborately designed visualizations to explore different modality data, offer diagnostic interpretive hints based on the constructed model, and enable comparative analyses of specific patients. Our approach is validated through two case studies and expert interviews, demonstrating its effectiveness in enhancing medical training.
} 


\graphicspath{{figs/}{figures/}{pictures/}{images/}{./}} 

\usepackage{tabu}                      
\usepackage{booktabs}                  
\usepackage{lipsum}                    
\usepackage{mwe}                       
\usepackage{mathptmx}                  
\usepackage{balance}
\usepackage{microtype}                 
\PassOptionsToPackage{warn}{textcomp}  
\usepackage{textcomp}                  
\usepackage{mathptmx}                  
\usepackage{times}                     
\usepackage{cite}                      

\usepackage{wrapfig}

\newcommand*\casebox[1]{\tikz[baseline=(char.base)]{
            \node[shape=rectangle,draw=white!30!black,fill=gray!15!white,inner sep=1pt, line width=0.5pt, text=black, font=\footnotesize] (char) {#1};}}
            
\usepackage{algorithm}
\usepackage{algorithmic}
\usepackage{paralist}
\usepackage{multirow}
\usepackage{fontawesome}

\begin{document}


\firstsection{Introduction}
\maketitle

\par Hospitals have a significant responsibility in training physicians by providing hands-on training and supervision from experienced physicians. However, given the workload of experienced physicians, it may be impractical for them to provide constant supervision and guidance during every diagnostic encounter. Additionally, it may be challenging for interns and novice physicians to assume full responsibility for diagnostic activities without adequate preparation~\cite{burgess2018mentorship}. To address these challenges, there is a need to transition from the conventional \textit{``see one, do one, and teach one''} mode of medical training to a \textit{``see one, practice many, and do one''} approach~\cite{vozenilek2004see}. This shift may offer a more effective and practical mode of training, enabling interns and novice physicians to gain valuable hands-on experience and develop their diagnostic capabilities.

\par To facilitate additional practice opportunities for novice physicians and interns without incurring a significant medical burden, Simulation-based Medical Education (SBME) has been proposed~\cite{al2010simulation,mcgaghie2010critical}. SBME is an educational method that employs simulation tools to emulate clinical scenarios, which is an integral component of medical training as it allows trainees to refine their diagnostic abilities at a reasonable expense~\cite{seropian2004simulation,srinivasan2006assessment}. Nonetheless, SBME evaluations are often immediate and aim to enhance learners' medical skills and practical expertise, rather than to \textbf{enhance the summative medical diagnosis} that emerges from time and experience~\cite{mcgaghie2010critical}. Furthermore, in practice, physicians must deal with health data from various sources, such as laboratory test results, radiological images, and medical texts, to diagnose a patient's condition. The multimodal nature of this health data presents additional challenges,  as \textbf{novices and interns may tend to overlook or excessively rely on certain modalities due to their lack of mindset and clinical experience.} This could lead to an incomplete understanding of the patient's data, which could compromise the effectiveness of the learning outcomes. To address this challenge and help novices and interns learn to better utilize multimodal medical data for decision-making, machine learning (ML) techniques have been employed to model data, make predictions, and serve as additional references for teaching purposes~\cite{shi2017multimodal,chen2019multi}. Despite this, the current literature~\cite{solares2020deep,piccialli2021survey} suggests that existing works primarily focus on adapting state-of-the-art techniques to improve model performance~\cite{ashuach2021multivi,faris2021intelligent,muhammad2021comprehensive,acosta2022multimodal} but \textbf{overlook the impact of individual modalities on the contribution to model results.} This can make it difficult for learners to comprehend the model's behavior and gain useful medical knowledge from it.

\par In this study, we aim to improve the simulation-based medical training workflow by designing interactions between interns and novice physicians and multimodal ML models trained from existing medical databases. To achieve this objective, we first conducted an observational study with four physicians with different levels of medical expertise and one data scientist to comprehend their primary needs and concerns regarding their mentoring and learning experiences. Based on the learning cycle~\cite{kolb1984experience,zigmont2011theoretical}, we summarize the learning process of the participants in three stages: clarifying the data situation and diagnostic tasks (i.e., learning goals), observing and comprehending each modality data (i.e., self-awareness), and reflecting on the analysis processes (i.e., self-adjustment). Furthermore, discussions were held with data science experts working in the hospital to discuss the application of multimodal ML models in medicine. A multimodal model was adapted for a showcased diagnostic task based on the multimodal data provided by the experts. Additionally, we propose a visual analytics system named \textit{DiagnosisAssistant} to assist interns and novice physicians enhance their learning experience. The system is built upon the complementary multimodal model described above and presents the joint impact of various modalities with intuitive visual representations to enable domain experts to better understand the model decision. The system also supports users in exploring various aspects of multimodal data and conducting comparative analyses for individual patients. The efficiency and reliability of the approach were validated through two case studies and expert interviews. This study makes the following contributions:
\begin{compactitem}
    \item We shadow and gain insight into the ``mentor-apprentice'' processes between experienced physicians and interns/novices.
    \item We facilitate and enhance the learning experience of interns and novice physicians by developing a visual analytics system embedded with a multimodal model.
    \item We demonstrate the validity and reliability of our approach through two case studies and expert interviews.
\end{compactitem}

\section{Related Work}
\par The related literature can be categorized into four distinct areas that overlap with this work: \textit{simulation-based medical education}, \textit{medical diagnosis with multimodal models}, \textit{post-hoc explainability techniques}, and \textit{medical data visualization}.

\subsection{Simulation-based Medical Education}
\par Simulation technology has become central to medical education, and the effectiveness of simulation-based learning depends on the type of simulators used. Simulators can be classified into three categories: \textit{part-task trainers}, \textit{computer-based systems}, and \textit{integrated simulators}~\cite{al2010simulation}. Part-task trainers cover basic procedural skills as well as high-fidelity virtual reality trainers with haptic feedback for complex medical procedures like surgery~\cite{hoopes2020home,kazan2016novel,tanya2020development}. Computer-based systems have the potential to create simulated patients or environments and provide interfaces for learners to interact with basic science material and receive feedback~\cite{chengoden2023metaverse,rooney2018simulation}. Integrated simulators combine realistic manikins with computer-driven features to simulate medical procedures and interactions~\cite{mcgaghie2016revisiting}. There are two types of integrated simulators: model-driven simulators (e.g., Human Patient Simulator) and instructor-driven simulators (e.g., "Noelle" Obstetrics Simulator)~\cite{al2016barriers,lutgendorf2017multidisciplinary}.

\par The majority of the cases mentioned above involve direct feedback from SBME, which focuses on enhancing learners' medical skills and operational proficiency rather than improving summative medical diagnosis and learning experiences. This study falls under the purview of computer-based systems, which aid interns and novice physicians in enhancing their learning experiences by constructing and revealing a multimodal model integrated into a visual analytics system. Furthermore, the system we have developed is designed to foster self-directed learning by emphasizing the learning cycle process~\cite{kolb1984experience,zigmont2011theoretical}. This approach allows users to enhance their knowledge and skills in a self-paced manner, enabling them to engage in a continual process of learning and personal development.

\subsection{Medical Diagnosis with Multimodal Models}
\label{sec:2.2}
\par Medical diagnosis involves utilizing heterogeneous data, including clinical records, laboratory tests, and radiological images. Multimodal models, driven by ML technologies, have emerged to assist in medical diagnosis by leveraging this diverse data. These models can be classified into two types based on their input fusion strategy: \textit{decision-level fusion} and \textit{feature-level fusion}~\cite{shi2017multimodal,li2019longitudinal,huang2020fusion} models. Decision-level fusion models integrate probabilistic or categorical predictions from unimodal models using techniques like averaging, weighted voting, or majority voting to generate a final multimodal prediction~\cite{holste2021end,kawahara2019seven,wang2021modeling}. For instance, Huang et al.~\cite{huang2020multimodal} used \textit{PENet} and a feedforward neural network on CT scans and electronic medical records (EMR) to detect pulmonary embolism, while Kawahara et al.~\cite{kawahara2018seven} employed convolutional neural networks (CNNs) on camera images and metadata for diagnosing skin lesions. However, decision-level fusion may lack interaction among hidden features despite being able to handle missing patterns. On the other hand, feature-level fusion methods extract raw data or multimodal features into a concise and informative representation for final prediction~\cite{shi2017multimodal,li2019longitudinal,huang2020fusion}. For example, Shi et al.~\cite{shi2017multimodal} utilized autoencoders to extract features from magnetic resonance images (MRI) and positron emission tomography (PET) for diagnosing Alzheimer's disease, while Chen et al.~\cite{chen2019multi} proposed an Attention Mutual-Enhance (AME) module to fuse features during the extraction phase for diagnosing cervical spondylosis. Feature-level fusion methods, compared to decision-level fusion, are generally better equipped to capture complex relationships between features from different modalities.

\par While numerous multimodal models employ \textit{feature-level fusion} techniques to enhance performance in specific tasks, their complex and hybrid architectures pose a significant challenge in understanding the internal workings of these models~\cite{muhammad2021comprehensive}. In contrast to previous approaches, our study takes a different direction by focusing on comparing the effectiveness of various fusion strategies within the context of a diagnostic task. Specifically, we have implemented a \textit{decision-level fusion} model that outperforms the feature-level model in terms of performance. Our primary research objective is to investigate the efficacy of fusion strategies, with a particular emphasis on the decision-level fusion approach. It is important to note that the interpretability of feature-level fusion models is beyond the scope of this study.

\subsection{Post-hoc Explainability Techniques}
\par Post-hoc interpretability involves revealing the inner workings of an ML model after its creation, allowing it to be applied to existing models. Interpretable methods can be classified as either model-specific or model-agnostic, depending on their compatibility with different models~\cite{arrieta2020explainable,payrovnaziri2020explainable}. Model-specific methods generate interpretations tailored to specific models, while model-agnostic methods can be applied to any model. This study focuses on model-agnostic approaches, specifically, \textit{surrogate model-based} and \textit{feature contribution-based} methods. Surrogate model-based approaches transform complex models into interpretable approximations, such as linear models~\cite{ribeiro2016should}, tree-based models~\cite{bastani2017interpretability}, or rule-based models~\cite{konig2008g,ribeiro2018anchors}, to mimic predictions of the original ``black box'' model~\cite{krause2016interacting,ribeiro2016should,guo2018lemna}. Feature contribution-based approaches investigate the impact of features on model decisions, such as permutation importance and SHAP, which measure global and local feature importance, respectively~\cite{henelius2014peek,lundberg2017unified,breiman2001random}. Attention-based neural networks have gained popularity but often lack clear explanations for their predictions~\cite{payrovnaziri2020explainable,choi2016retain,kwon2018retainvis}.

\par In our study, we incorporate various interpretability methods to analyze models derived from different modalities. Specifically, when dealing with the medical image modality, it is essential to consider its unique characteristics for accurate model interpretation. To address this, specific tools have been developed to accommodate these distinct characteristics~\cite{huff2021interpretation}. Within this context, we utilize the Guided Grad-CAM approach~\cite{selvaraju2017grad}, which has demonstrated its value in multi-class settings due to its class-specific nature~\cite{huff2021interpretation}. For the text modality, we employ the interpretability of transformers approach~\cite{chefer2021transformer}. Lastly, for the indicator modality, SHAP has been utilized. By incorporating multiple techniques, we acknowledge the ability of each approach to effectively highlight its respective strengths and provide insights into the transformation of different types of raw data processed through diverse models. Our aim is to enhance the learning experience of interns and novice physicians in medical diagnostics by utilizing multimodal modeling and visualizing historical medical records as a proxy.

\subsection{Medical Data Visualization}
\par Medical data holds valuable information and vast research potential. However, the complex nature of electronic health records, sourced from diverse origins with heterogeneous and temporal variations~\cite{dinov2016predictive}, makes it challenging to identify underlying patterns. Medical data visualization empowers researchers, professionals, and even patients to explore, examine, and make informed decisions about patient health~\cite{caban2015visual}. This discussion focuses on visualizing two primary medical data types: time series and cohorts. Time series encompass physiological signal data (e.g., heart rate, respiratory rate, oxygen saturation) and patient-centered event series (e.g., treatment and medication records)~\cite{guo2021survey}. Timeline-based visualizations, such as point or interval plots, provide detailed information about physiological data~\cite{shahar2006distributed}. Recent advancements include \textit{ThreadState} proposed by Wang et al.~\cite{wang2021threadstates}, which employs a new glyph matrix design and Sankey plots to identify disease progression states. Other approaches involve summarizing patient history through timelines~\cite{shahar2006distributed} and storylines~\cite{baumgartl2020search}. Retrospective cohort studies~\cite{kuiper2015social} and visual analysis of longitudinal cohort data~\cite{bernard2018using} have been valuable for various purposes, such as assessing healthcare team performance~\cite{malik2015cohort}, summarizing disease states ~\cite{wang2021threadstates}, and comparing medical image attributes~\cite{calisto2021introduction}.

\par Several studies have explored multimodal medical data visualization. For example, Raidou et al.~\cite{raidou2018bladder} integrated multiple visualization techniques, including timeline, scatterplot, heatmap, brushing and linking, and interactive filters, to explore radiation-induced bladder toxicity in a cohort study using clinical data such as treatment events and patient demographics. M{\"o}rth et al.~\cite{morth2020radex} integrated clinical and radiological data, employing heatmaps, boxplots, scatterplots, and decision trees to facilitate the exploration of multiparametric studies for radiomic tumor profiling. Sugathan et al.~\cite{sugathan2022longitudinal} introduced a longitudinal visualization approach for examining multiple sclerosis lesions over time, enabling both qualitative and quantitative analysis of lesion progression.

\par Unlike previous studies that primarily relied on specialized visualizations for depicting multimodal data, our approach stands out by providing visual explanations that comprehensively explore the influence of multimodal characteristics. Moreover, our research goes a step further by creating visualizations encompassing various data modalities to assist interns and novice physicians in effectively navigating medical data, spanning from patient cohorts to individual patients.

\section{Observational Study}
\label{observational_study}
\subsection{About the Team and the Mentoring Process}
\par To gain a deeper understanding of  customary procedures within the medical diagnostic process, as well as the learning experiences of novice physicians and interns, we collaborated with a team of five domain experts from a reputable local hospital. The team includes two interns (\textbf{I1}, male; \textbf{I2}, female), two chief physicians (\textbf{E3}, male; \textbf{E4}, female), and one data scientist (\textbf{E5}, male) responsible for medical data engineering within the hospital. 
Notably, \textbf{E3} served as an experienced physician and mentor to both \textbf{I1} and \textbf{I2}.

\par In the medical field, the relationship between novice physicians, interns, and their supervisors follows a customary ``mentor-apprentice'' dynamic. Novice physicians and interns are always under the direct supervision of experienced residents or attending physicians. The medical prescriptions written by interns and novice physicians undergo verification and endorsement by their supervising physicians, including documentation of the courses of action they have proposed. Ultimately, the attending physician assumes responsibility for the patient's care and oversees the decisions made by the interns and novice physicians. The interns and novices are required to report their analyses to their supervising physicians, who engage in comprehensive discussions based on the provided information and the best available evidence. The process of diagnosing a patient from multiple data modalities typically follows a top-down approach. Initially, patient information is gathered from various sources, each offering a different perspective. Physicians then review the data from all modalities to gain an overview and carefully examine any abnormalities detected in each modality. Based on these observations, physicians form potential diagnoses for each data modality. In essence, information is shared, and the control over diagnosis and treatment becomes more comprehensive in theory.

\subsection{Experts' Concerns and Bottlenecks}
\par Despite receiving standard medical training, interns and novice physicians encounter various challenges during the practical learning process, leading to potential issues.
\par \textbf{Communication barriers may arise} due to a shortage of doctors in the healthcare system, particularly in primary and county hospitals, for various reasons. ``\textit{In certain situations, interns and inexperienced physicians may not have sufficient time or opportunities to thoroughly discuss the medical diagnostic decision-making process with their colleagues,}'' said \textbf{E3}. This lack of communication may lead to errors and hurt patient outcomes. \textbf{I1} further highlighted that in real diagnostic scenarios, interns might refrain from asking questions, which could also impede their learning process. ``\textit{Sometimes I want to ask a question, but given that this is a real medical diagnostic scenario, I choose not to ask the question,}'' said \textbf{I1}.
\par \textbf{Experienced physicians and interns/novices exhibit distinct mindsets, and relying solely on one's experience can be a ``double-edged sword''.} Interns/novices with limited experience are more inclined to consult textbooks and embrace challenges. As recounted by \textbf{I1}, an experienced physician had initially diagnosed a patient with asthma, but the patient's condition did not improve despite multiple treatment attempts. However, an intern scrutinized the patient's physical examination data and identified the trigeminal sign (i.e., depressions in the supraclavicular fossa, and intercostal space during inspiration), indicating the possibility of airway stenosis. The intern's observations were confirmed by their supervisor, who then ordered pulmonary function tests, leading to a correct diagnosis. This case underscores the significance of obtaining patient information from various sources and recognizing potential links between different modalities in order to improve overall diagnostic accuracy.
\par \textbf{Interns and novice physicians often lack opportunities to practice their skills on actual patients.} \textbf{E4} conducted a survey of $11$ patients, and while four were willing to be seen by an intern for minor illnesses, seven were apprehensive and expressed concerns about misdiagnosis due to the intern's lack of experience, ``\textit{reluctance because the intern is too young and inexperienced and afraid of misdiagnosis.}'' However, there were also patients who recognized the importance of interns having opportunities to gain clinical experience. \textbf{E3} acknowledged that while patients have the right to refuse care from interns, ``\textit{if most patients refuse, it could hinder the growth and development of interns, leading to a discontinuity in the quality of doctors in the future}''. To streamline their learning process, a common approach in SBME is a standardized test~\cite{issenberg2011setting}. However, such tests may limit learners' critical and creative thinking abilities as they focus on reference answers. For example, \textit{``how would a physician judge if there were only two specific data modalities''} and \textit{``whether a different diagnosis would occur when certain data modalities change.''} These issues frequently arise in real clinical scenarios where physicians must often make judgments without all necessary data modalities simultaneously available. Thus, there is a pressing need to develop a medical diagnostic platform to enhance the learning experiences of interns and novice physicians.

\begin{figure*}
\centering
\vspace{-2mm}
\includegraphics[width=\textwidth]{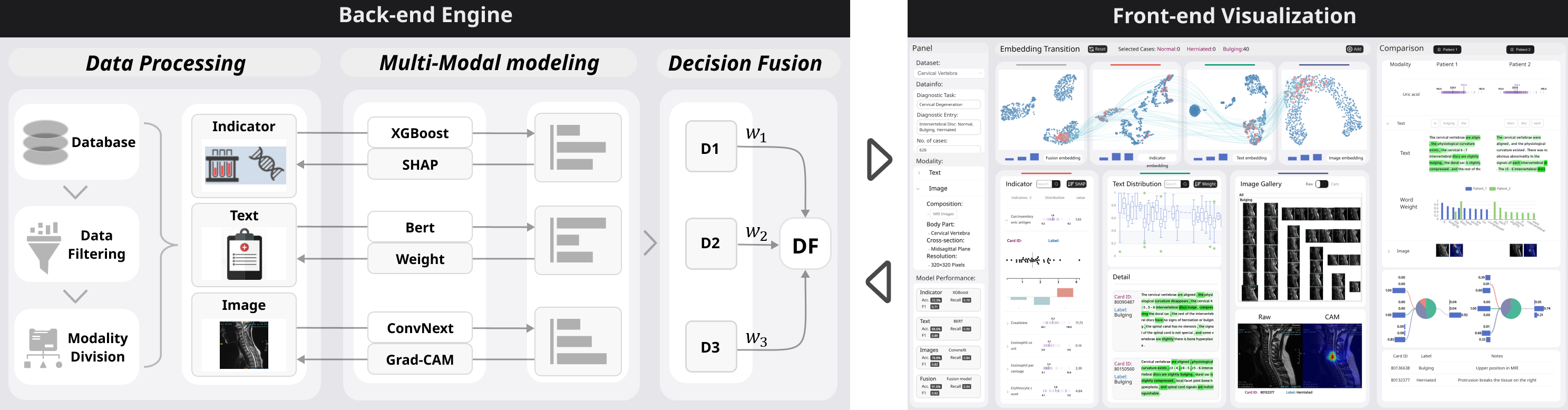}
\vspace{-6mm}
\caption{The processing of medical data in different forms is conducted by the \textbf{back-end engine}. It uses a multimodal model to make predictions and employs interpretability techniques to explain them. The \textbf{front-end visualization} facilitates interactive exploration for improved diagnostic learning.}
\label{fig:pipeline}
\vspace{-7mm}
\end{figure*}

\subsection{Experts' Needs and Expectations}
\par Following interviews with all experts, we compiled a list of requirements to improve the medical diagnostic learning experience for interns and novices by addressing potential obstacles and concerns. Our system is intended to be consistent with the learning cycle principles, allowing users to achieve their learning objectives, enhance self-awareness, and engage in self-adjustment. First, \underline{learning objectives} for our system should comprise of clarifying medical diagnostic tasks and comprehending the general data situation \textbf{(R.1)}. Second, to develop \underline{self-awareness}, visual cues for each modality data and simple observation and comprehension of data are necessary \textbf{(R.2 -- R.5)}. Third, it is crucial to encourage users to \underline{reflect on their analysis processes} and compare their results to other case studies to bolster their learning experience and deepen their understanding of the diagnostic task \textbf{(R.6)}.

\par \textbf{R.1 Describe the focal medical diagnostic tasks and present data statistics.} According to the experts, it is essential to provide statistics regarding the multimodal dataset, including data quality and sources for each modality. Additionally, a clear definition of medical diagnostic tasks should be provided.

\par \textbf{R.2 Develop a reliable and interpretable ML model that can capture the diagnosis process.} According to the experts, ML techniques, especially deep learning methods, are highly effective in this regard. Data scientist \textbf{E5} affirmed that their prior modeling experiments have also yielded satisfactory outcomes. However, they emphasized that certain aspects need to be addressed if advanced models are to be implemented in actual diagnostic scenarios. Specifically, the model's accuracy must be sufficiently high, and it must also capture the diagnostic process accurately. \textbf{E5} further stated that ``\textit{it is crucial to understand how the model arrived at its conclusions and whether its decision-making process aligns with medical findings}''.

\par \textbf{R.3 Convert the diagnostic process into a user-friendly representation.} Once a reliable ML model has been developed, the focus shifts toward creating an intuitive representation of the diagnostic process for historical cases. The experts highlighted the importance of demonstrating the model's functionality in a clear and easily understandable manner. As \textbf{E3} noted, conveying the abstract experience in a user-friendly manner would be beneficial for all involved.

\par \textbf{R.4 Show the performance of the model and the contribution of each data modality.} Although inexperienced physicians and trainees have the potential to develop a broader perspective that transcends their specialization, they may fall into the trap of over-reliance on a single modality due to their limited clinical exposure. \textbf{E3} provided a case study where the negligence of clinical cues and excessive reliance on CT imaging resulted in delayed detection of bowel cancer in a patient for a considerable length of time. Furthermore, as \textbf{E4} highlighted, ``\textit{imaging techniques may not capture all the relevant details, and their accuracy may be affected by factors such as the angle and method of capture}''. Hence, it is essential for novice physicians and interns to comprehend the performance of the model and the contribution of each modality to the final diagnosis.

\par \textbf{R.5 Reveal the relationship between different modalities.} When only one type of data is available, inexperienced medical professionals can easily make assessments based on that single modality. However, when faced with data from multiple modalities, they may encounter difficulties and confusion. According to \textbf{E2}, this difficulty arises from a lack of proficiency in correlating information across different modalities. It is crucial for them to establish connections between these modalities, considering that some modalities may present contradictory findings. For example, while a clinically recommended modality may indicate the presence of \textit{Benign Prostatic Hyperplasia}, no abnormalities may be detected through a medical ultrasound. In such cases, a comprehensive patient analysis, including factors like age and relevant symptoms, should guide the diagnosis to favor the ultrasound results. It is important to note that different diseases may require different interpretations of the relationships between modalities.

\par \textbf{R.6 Support comparative analysis of individual patients and maintain data provenance.} Conducting comparative analysis on typical cases of specific diseases holds significant importance in the realm of clinical practice. ``\textit{By comparing individual patients from diverse groups, we can broaden our knowledge and comprehension of varied disease pathologies and patient cohorts}'', said \textbf{I2}. This aids interns and novice practitioners in developing their diagnostic skills and enables them to make accurate diagnoses in real-life scenarios. Additionally, the system should maintain a log of actions to monitor and track the comparison process. In summary, the comparative analysis of individual patient cases plays a crucial role in enhancing diagnostic accuracy and minimizing the risk of misdiagnosis and missed diagnoses.

\section{DiagnosisAssistant}
\par Based on the identified needs and requirements, we have developed a novel visual analytics system, namely \textit{DiagnosisAssistant}, to enhance the diagnostic learning experience of interns and novice physicians. The system's architecture (\cref{fig:pipeline}) encompasses a back-end engine responsible for processing medical data from three distinct modalities, namely \textit{indicator}, \textit{text}, and \textit{image}. The processed data is then fed into a reliable multimodal model that generates predictions by leveraging the aforementioned modalities. To promote transparency, innovative interpretability techniques are employed to elucidate the model's predictions. The front-end visualization empowers users to interactively explore the processed data across the three modalities, thereby enhancing the diagnostic learning experience for interns and novice physicians.

\subsection{Back-end Engine}
\subsubsection{Data}
\begin{figure*}[h]
\centering
\vspace{-1mm}
\includegraphics[width=\textwidth]{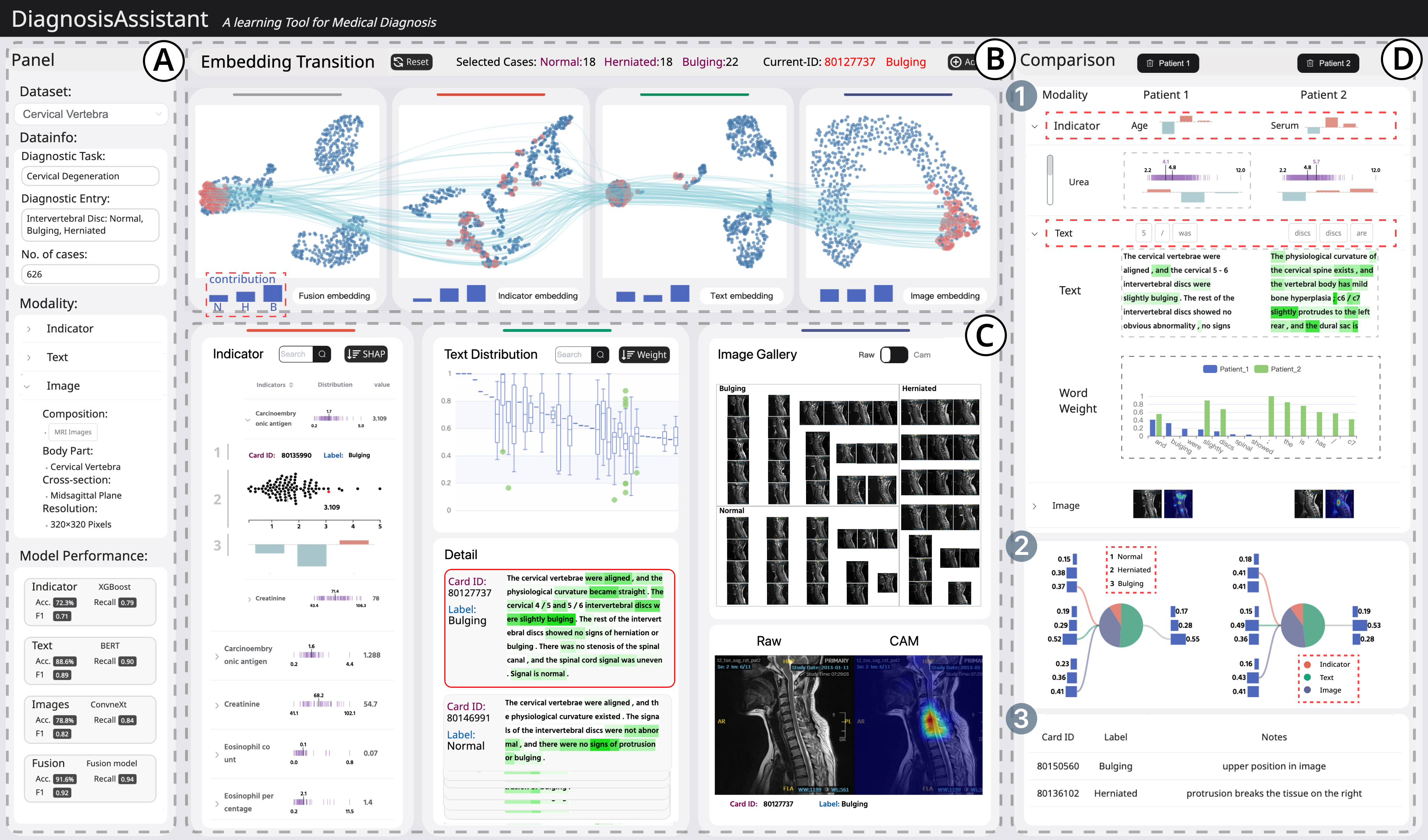}
\vspace{-6mm}
\caption{The system interface of \textit{DiagnosisAssistant} contains (A) the User panel, (B) the Embedding Transition View, (C) the Modality Exploration View, and (D) the Comparison View. }
\label{fig:system}
\vspace{-6mm}
\end{figure*}

\par \textit{DiagnosisAssistant}, is designed to process heterogeneous clinical data, construct robust models, and provide diagnostic insights based on interpretability techniques. To evaluate the system's performance, we collaborate closely with physicians from a prominent local hospital and used a real-life clinical dataset for diagnosing cervical spine disorders. The dataset comprises $750$ patient records collected during hospital visits for cervical spine discomfort between 2012 and 2013. The patients' ages range from $21$ to $82$ years, with a male-to-female ratio of $1.16:1$. Each patient has a unique CardID that corresponds to a set of clinical records, including \textit{demographic information}, \textit{laboratory test results}, \textit{magnetic resonance imaging (MRI) images}, \textit{clinical reports}, and \textit{diagnostic findings}. We organize the data into three modalities, namely, the \textit{indicator modality}, \textit{text modality}, and \textit{image modality}. The indicator modality consists of laboratory test results and demographic information, including gender, age, height, and weight. The unstructured data includes clinical reports and MRI images. Prior to model training, the dataset was carefully preprocessed and checked, resulting in 626 retained instances.

\subsubsection{Multimodal Modeling and Unfolding}
\par In this subsection, we first present our approach for developing a multimodal model and subsequently outline our method for unfolding the diagnostic insights derived from the constructed model. Specifically, the design of our model revolves around the classification of patients into three distinct categories, namely, \textit{normal, herniated, bulging}, as suggested by domain experts.

\begin{table}[h]
\centering
\vspace{-3mm}
\caption{Model Performance.}
\vspace{-3mm}
\label{tab:models}
\resizebox{0.80\columnwidth}{!}{%
\begin{tabular}{ccccc}
\hline
\textbf{Modality}          & \textbf{Model}   & \textbf{Acc.}   & \textbf{Recall} & \textbf{F1}   \\ \hline
\multirow{2}{*}{Indicator} & XGBoost          & \textbf{72.3\%} & \textbf{0.79}   & \textbf{0.75} \\ \cline{2-5} 
                           & Random Forest    & 69.5\%          & 0.71            & 0.70          \\ \hline
\multirow{2}{*}{Text}      & ClinicalBERT     & 88.6\%          & \textbf{0.90}   & \textbf{0.89} \\ \cline{2-5} 
                           & BioBERT          & \textbf{89.1\%} & 0.85            & 0.87          \\ \hline
\multirow{2}{*}{Image}     & ConvNeXt         & \textbf{78.8\%} & \textbf{0.84}   & \textbf{0.82} \\ \cline{2-5} 
                           & Swin Transformer & 75.2\%          & 0.79            & 0.77          \\ \hline
\end{tabular}%
}
\vspace{-1mm}
\end{table}

\par We explore various machine learning techniques for modeling the data across multiple modalities, as summarized in \cref{tab:models}. Subsequently, we select three models for each modality, \textit{XGBoost}~\cite{chen2016xgboost} for indicator data, \textit{ClinicalBERT}~\cite{alsentzer-etal-2019-publicly} for textural data, and \textit{ConvNeXt}~\cite{liu2022convnet} for image data, owing to their superior performance. We fine-tune the hyper-parameters of the three models using grid search~\cite{ensor1997stochastic} with k-fold cross-validation~\cite{stone1974cross}. The dataset is divided into training and validation sets using a random selection method with a ratio of 75:25 for training and validation, respectively. \Cref{tab:models} lists the performance of the models on the validation set with the optimal hyper-parameter settings.

\par \textbf{Fusion Strategy.} Fusing heterogeneous information from multimodal data is a common strategy to improve model performance~\cite{cui2022deep}. In this study, we investigate two fusion strategies: \textit{decision-level} fusion and \textit{feature-level} fusion (discussed in \autoref{sec:2.2}). In \textbf{feature-level fusion}, we concatenate the raw indicator data with the output from the penultimate layer in the \textit{ConvNeXt} and \textit{ClinicalBERT} models and then feed the concatenated features into a $12$-head, $12$-layer transformer. In \textbf{decision-level fusion}, we adopt a weighted voting strategy and employ multiclass perception to learn the weights of each modality. As shown in \cref{tab:fusion}, the two strategies perform similarly. However, the decision-level fusion strategy aligns better with physicians' diagnostic process and is more robust and scalable in the absence of modalities. Therefore we select the decision-level fusion strategy for our model.

\begin{table}[h]
 \vspace{-3mm}
\centering
\caption{Experimental Results of Fusion Strategy.}
\vspace{-3mm}
\label{tab:fusion}
\footnotesize
\resizebox{0.6\columnwidth}{!}{%
\renewcommand\arraystretch{1.4}
\begin{tabular}{cccc}
\hline
\textbf{Strategy}      & \textbf{Acc.}    & \textbf{Recall}  & \textbf{F1}   \\ \hline
Feature-level  & \textbf{92.1\%} & 0.92   & 0.92 \\ \hline
Decision-level & 91.6\% & \textbf{0.94}   & 0.92 \\ \hline
\end{tabular}%
}
\vspace{-3mm}
\end{table}

\par \textbf{Interpretability Towards Diagnosis.} To improve users' learning experience, we apply multiple post-hoc interpretability techniques. Specifically, \textit{Guided Grad-CAM}~\cite{selvaraju2017grad} is utilized to generate saliency maps on each MRI image and highlight important areas. The method proposed by Chefer et al.~\cite{chefer2021transformer} is adopted to interpret the transformer model and highlight important words. For the indicator-modal model, \textit{SHAP} is used to determine the contribution of each feature to a particular decision. By providing saliency maps, key text highlighting, and feature contribution quantification, comprehensive hints about the diagnostic focus of each case are given to the users to aid in their diagnostic learning process. All these explanations are combined to offer interpretability to the model's decisions.

\subsubsection{Embedding Generation}
\par Our objective is to provide users with an understanding of the data distribution and model behaviors for each modality by extracting and visually summarizing the data embeddings. For the indicator data, we directly utilize the raw data without additional processing due to its low dimensionality. In the case of \textit{ClinicalBERT}, which follows the architecture of the \textit{BERT} model and comprises 12 transformer layers, each layer generates 768-dimensional embeddings that capture semantic features at various levels. To retain the maximum amount of valid information, we employ common techniques such as summing, averaging, and concatenating selected or all layer embeddings. We adopt the bit-wise sum of all token embeddings as the final 768-dimensional text representation, striking a balance between computational efficiency and information retention. As for \textit{ConvNeXt}, an optimized CNN that follows the classic CNN architecture, we extract the output from the penultimate layer (i.e., the input of the classifier layer) to obtain a 768-dimensional image representation. To encompass the model's overall understanding of the data distribution, we define a fusion embedding by concatenating the three embedding vectors and weighting each vector element according to its modality.

\subsection{Front-end Visualization}
\par To facilitate the exploration and comprehension of multimodal medical data by junior physicians, we have developed the \textit{DiagnosisAssistant} interface (\cref{fig:system}), which comprises four distinct views. The \textit{User Panel} offers users an overview of the multimodal medical dataset and model performance, while also providing descriptive statistics about the dataset, data quality, and sources for each modality (\textbf{R.1, R.3}). The \textit{Embedding Transition View} (\textbf{R.2, R.4, R.5}) and the \textit{Modality Exploration View} (\textbf{R.3, R.4, R.5}) assist users in identifying and comparing different patient cohorts across modalities. Lastly, the \textit{Comparison View} allows for the comparative analysis of individual patients, aiding in the improved understanding of specific diseases (\textbf{R.6}).

\subsubsection{Embedding Transition View}
\par To examine the patient collections' features in various embedding spaces, an embedding transition view (\cref{fig:system}(B)) has been developed. This view connects the fusion model embedding and the three modality data (i.e., indicators, text, images) embedding spaces \textbf{(R.2)}. Each patient is represented by a node, and the connections between each projection allow users to trace the same group of patients across different embeddings. To generate two-dimensional projections, the \textit{t-SNE} method~\cite{van2008visualizing} is chosen as it``\textit{reveals meaningful insights about the data and shows superiority in generating two-dimensional projection}~\cite{li2018embeddingvis}.'' The users can use the lasso tool to select a group of patients, and the system will display the distribution of the selected patients by class. Moreover, each modality data for each patient contributes to the final model output, where the contribution is a probability estimate for each category in the resulting model output (\textbf{R.4}, \textbf{R.5}). To show each modality data's contributions to different prediction classes (three classes in the running example), bar charts are used. The contribution value is determined by the \textit{probability mass function} (PMF)~\cite{stewart2009probability}, which characterizes the distribution of discrete random variables. The contribution values attached to the fusion embedding projection are calculated as weighted sums of the three modalities.

\subsubsection{Modality Exploration View} 
\par As each modality possesses its distinct characteristics, we propose visual interfaces and interaction for each modality individually to present intuitive representations \textbf{(R.3, R.5)}.

\begin{figure}[h]
\centering
\vspace{-3mm}
\includegraphics[width=\columnwidth]{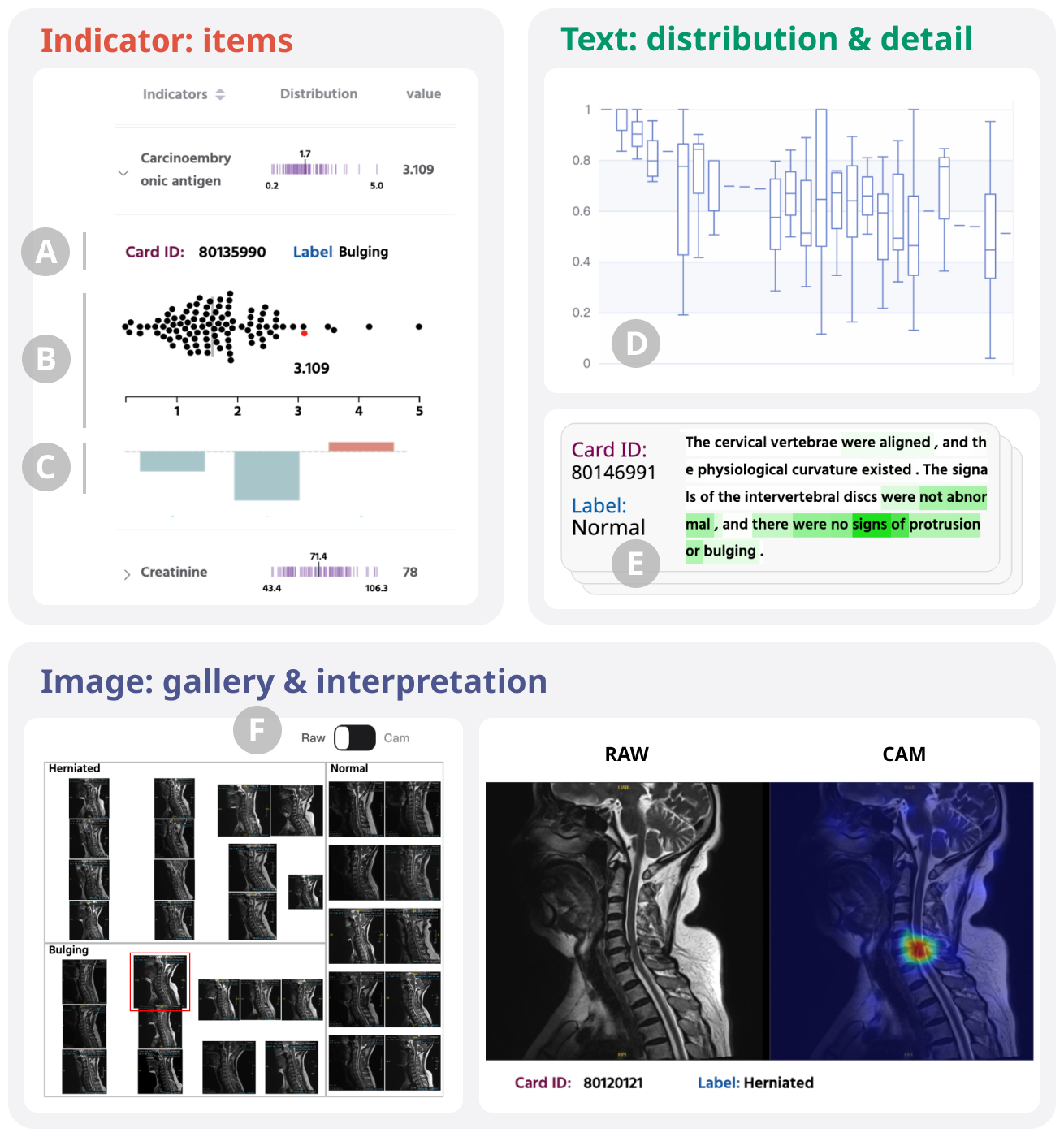}
\vspace{-6mm}
\caption{The Modality Exploration View includes \textcolor[RGB]{216, 79, 61}{Indicator}, \textcolor[RGB]{23, 139, 96}{Text} and \textcolor[RGB]{68, 76, 128}{Image} modalities.}
\label{fig:modalities}
\vspace{-3mm}
\end{figure}

\par \textbf{Indicator Modality.} Based on the physicians' suggestions, we extract $34$ pertinent indicators \textbf{(R.1)}. These indicators are structured in a collapsible table, where each row corresponds to one indicator. In the \textit{collapsed mode}, each row includes the indicator name, an approximate distribution of indicators concerning the lassoed patient groups, and the mean value, thereby enabling the user to gain an overview of the indicator modality. The overall distribution of indicator data is exhibited using strip charts. The details can be viewed by clicking and expanding each feature, revealing a Beeswarm plot illustrating the data value of each instance \textbf{(R.6)}. When the user hovers over a point, the patient information (card-id and class) and the indicator value are displayed (\cref{fig:modalities}(A)(B)). If a point is clicked, it turns red, and the corresponding values of all the other indicators change accordingly. A bar chart is employed to exhibit the \textit{Shapely value} (\cref{fig:modalities}(C)), which initially represents the contribution of the selected patients to different classes. If a point in the beeswarm plot is clicked, it goes to the specific Shapely value, where the color encodes its contribution to the predicted value (red for positive and blue for negative). The table supports operations like sorting by contribution and filtering by indicators, providing clinicians with more control in the exploration \textbf{(R.3)}.

\par \textbf{Design Alternatives.} We have evaluated four types of visualizations, namely \textit{Histogram}, \textit{Density Plot}, \textit{Violin Plot}, and \textit{Beeswarm Plot}. It was observed that Density and Violin plots are appropriate for representing continuous data, but inadequate for discrete data. In contrast, histograms employ interval grouping and frequency counting to facilitate the visualization of data distribution. However, when it comes to depicting data density and clustering, Beeswarm Plots have exhibited distinct advantages over histograms~\cite{goedhart2021superplotsofdata}. Furthermore, Beeswarm Plots have been found to offer a more lucid and efficient representation of the indicator distribution. After consultation with experts, it was determined that the Beeswarm Plot was the most suitable option.

\par \textbf{Text Modality.} We use two representations to show the text modality. The top part represents a boxplot demonstrating the word weight of medical text for chosen patients. The words are presented in descending order based on their average weight, as depicted in \cref{fig:modalities}(D). The boxplot is a superior alternative to word clouds, as it can be aligned to enable a comparative analysis of word weights among diverse patient groups. Physicians can easily identify high-weighted terms, which can serve as visual cues for further exploration \textbf{(R.3)}. The lower part of the modality consists of blocks of medical text associated with the selected patients, arranged in ascending order according to their card ID. The words with higher weights are presented in a darker color, and the text block also displays the patient's card-id and class on the left side (\cref{fig:modalities}(E)). As the expert views the text, they can quickly identify some essential keywords, such as ``protrusion'' and ``bulging'' in the spine dataset, which physicians are particularly sensitive to.

\par \textbf{Image Modality.} The treemap is employed to categorize images based on various categories, facilitating a panoramic view of the gallery of selected patients. By clicking the button shown in \cref{fig:modalities}(F), the gallery can be switched between RAW image mode and CAM (Class Activation Mapping) mode. While browsing through the thumbnails in the gallery, if a user desires to examine a specific image, they can click on it to view the details on a larger scale. Both RAW and CAM information are displayed in parallel, providing users with a comprehensive understanding of the diagnostic focus of the image modality \textbf{(R.3)}.

\subsubsection{Comparison View}
\par For the purpose of clinical learning and reasoning, the \textit{Comparison View} facilitates a more detailed analysis of patient-specific similarities and differences \textbf{(R.6)}. Through this view, users can perform a more fine-grained comparison of individual patients, considering both aggregated multimodal data and detailed modality fusion. Additionally, users have the option to record notes for future reference.

\begin{figure}[h]
\centering
\vspace{-3mm}
\includegraphics[width=\columnwidth]{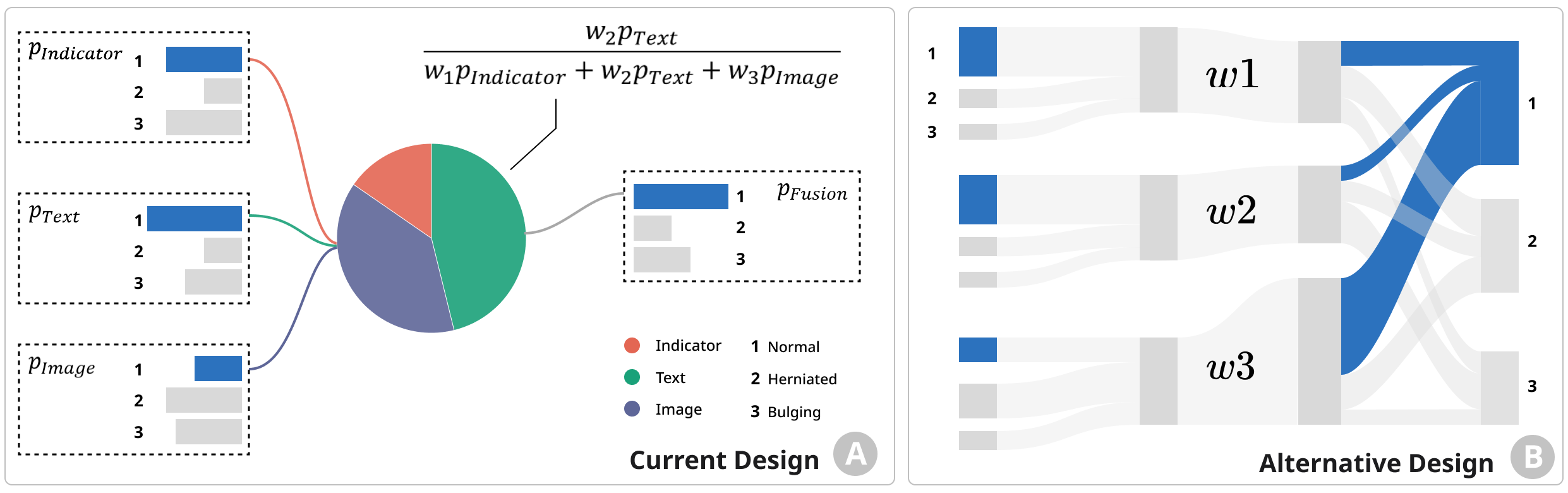}
\vspace{-6mm}
\caption{The Sankey diagram-based design (B) for probability fusion is less space-efficient and more cluttered compared to our design (A).}
\label{fig:prob_fusion}
\vspace{-3mm}
\end{figure}

\begin{figure*}[h]
\centering
\vspace{-2mm}
\includegraphics[width=\textwidth]{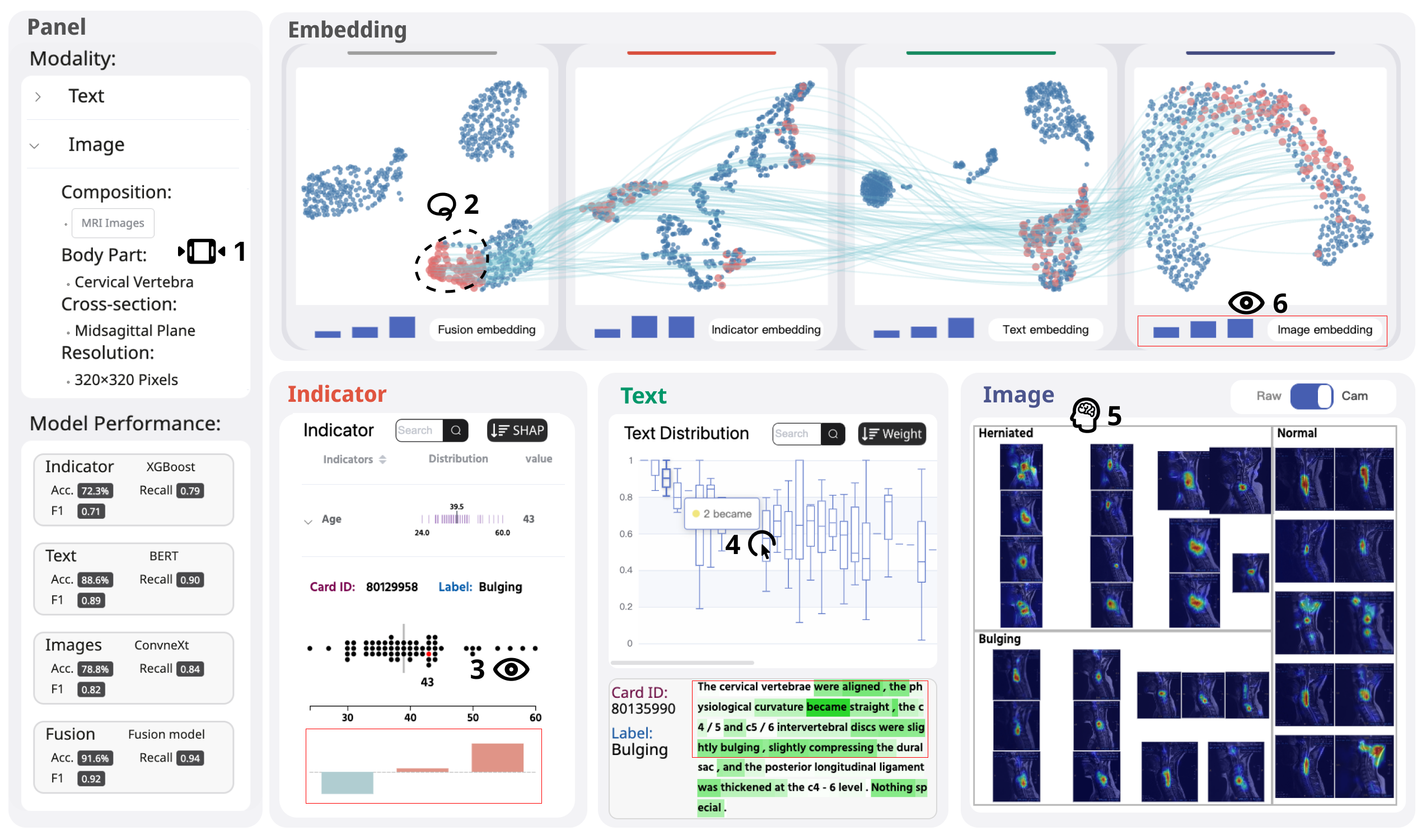}
\vspace{-6mm}
\caption{Case I: \textbf{E3} and \textbf{E4} conducted a retrospective analysis conducted to explore the multimodal data on cervical spine degeneration. \protect\casebox{1} They first understood the data situation and diagnostic tasks through the \textbf{User Panel}. \protect\casebox{2} Then, the experts utilized the \textit{Embedding Transition View} to identify patient groups and behavior across modalities. \protect\casebox{3-6} By thoroughly analyzing inter- and intra-modality, the experts obtained insightful results that were largely consistent with their expectations.}
\label{fig:case1}
\vspace{-6mm}
\end{figure*}

\par \textbf{Instance Comparison.} To present the diverse modalities of data, each with a unique structure for each patient, an aggregated approach is employed \textbf{(R.3)}. The multimodal information is organized into a collapsible table, as shown in \cref{fig:system}(D1). Each row within the table represents a single modality, while the three columns provide information about the modality types and the two selected patients. The first column lists the three modality types: indicators, text, and images. For each modality, the second and third columns display key information for each of the two patients \textbf{(R.5)}. This information includes indicator distribution and \textit{shapely value} data for the indicator modality, comparisons of medical text and their associated weights for the text modality, and RAW images alongside their CAM modes for the image modality. As a supplementary improvement, the information presented within the red box in \cref{fig:system}(D1) of the system denotes the feature that exhibits the highest SHAP value, alongside the top three weighted textual elements for each modality corresponding to individual patients. Moreover, upon expansion, a scrollable roster of information encompassing all features will be made visible. The visual encodings used in this table are consistent with those used in the \textit{Modality Exploration View}, ensuring a cohesive exploration experience for individual patients.

\par \textbf{Probability Fusion.} The process of decision fusion for the selected patients is depicted using hybrid visualizations (\cref{fig:prob_fusion}). On the left side, three groups of horizontal bars illustrate the predicted probability of the selected patient for each of the three modalities, listed from top to bottom. On the right side, a separate group of bars displays the final prediction after decision fusion has occurred. Each bar in a group corresponds to a diagnostic entry within the dataset. The central pie chart provides a visual representation of the percentage contribution of each modality to the fusion probability at the item level. Links connect the bars of corresponding entries in each group, while the pie and links are colored to match their source modality, facilitating the identification of modal correspondences. When hovering over a bar, the links connect to each bar of the current entry and depict the contribution of each modality to the fusion probability using the pie chart.

\par \textbf{Design Alternatives.} Initially, the Sankey diagram design was explored as a means of visualizing the decision fusion process (\cref{fig:prob_fusion}(B)). However, it was discovered that this design did not entirely align with the multi-modal fusion task at hand. Linking individual modality bins and final probability bins using Sankey links would have resulted in visual clutter. Therefore, experts were consulted, and it was decided that a more intuitive and easier-to-interpret display of the fusion results could be achieved through the use of simple bar charts and pie charts.

\par \textbf{Learning Recording.} To further augment the learning experience, the system provides users with the opportunity to document their reflections on a typical patient or a patient with an atypical diagnosis, thereby facilitating future reviews, as illustrated in \cref{fig:system}(D3). This feature is complemented by expert advice (\textbf{R.6}), and is expected to be of significant value. For instance, certain categories of cervical spine degeneration, such as ``herniation'' and ``bulge'', can be particularly perplexing to interns and novice physicians. Consequently, they may choose to document representative cases of each type and leverage the system's capabilities, alongside the guidance of experienced physicians, to systematically explore and learn from them.

\section{Evaluation}
\par We evaluated the effectiveness of our system through two case studies. Case I focused on experienced physicians, demonstrating how the \textit{DiagnosisAssistant} addressed their concerns and met their requirements. Case II explored whether our system improved the learning experience of interns and novice physicians in medical scenarios. Before conducting these case studies, we organized a tutorial session with the experts involved. They were introduced to the visualization designs, interactions, and workflow of our system. After a brief familiarization period of approximately $10$ minutes, the experts engaged in one-hour case studies while sharing their thoughts in a think-aloud manner. We then conducted interviews to gather their feedback on our approach.

\subsection{Case I: Retrospective Analysis}
\par We collaborated with \textbf{E3 -- E4}, who possess an average of $15$ years of experience in the relevant field, to perform a retrospective analysis. These physicians are well-versed in the cervical spine dataset and possess fundamental statistical knowledge pertaining to the various patient types across different data modalities. During the evaluation, the physicians utilized the \textit{DiagnosisAssistant} to explore the multimodal data concerning cervical degeneration and to assess whether our system could facilitate the learning process for junior physicians.

\par \textbf{Clarify the data situation and diagnostic tasks.} After loading the data, the experts first referred to the \textit{user panel} to understand the dataset and the model performances (\cref{fig:case1}(1)). 
\textbf{E4} asserted that ``\textit{the user panel gives me a general understanding of the data, a clear awareness of the diagnostic task at hand, and the model's performance.}''

\par \textbf{Target certain patient groups.} Next, the experts referred to the \textit{Embedding Transition View} to identify specific patient groups and to comprehend the behavior of certain patients under different modalities. By utilizing the lasso tool on each cluster, \textbf{E3} discovered that the clusters in the fusion embedding hold significance, where each cluster corresponds to a patient cohort with identical diagnostic outcomes (\cref{fig:case1}(2)). Furthermore, he focused on patients predicted as ``bulging''. By following the links, he observed that the patient group clustered more effectively in the image and text modalities, whereas the indicator modality resulted in smaller clusters.

\par \textbf{Analyze inter- and intra-modality.} ``\textit{The key for junior physicians to learn the diagnosis is to learn from the raw data of each modality, quickly locate abnormalities in each modality, and then integrate the information with medical knowledge,}'' said \textbf{E3}. Consequently, \textbf{E3} began to investigate specific modality data in the \textit{Modality Exploration View}. He discovered that \textit{``became''} plays a significant role in the majority of herniated patients, where most text descriptions possess phrases such as \textit{``physiological curvature became straight, the x/x intervertebral disc was slightly narrowed.''} (\cref{fig:case1}(4)). In the image modality, \textbf{E3} was pleased to note that the highlighted regions corresponded to his knowledge in making the judgment (\cref{fig:case1}(5)). Lastly, despite the subpar performance of the indicator modality, the indicator \textit{``Age''} was mainly distributed among individuals aged $24$-$50$ years, with a higher SHAP value, consistent with their observation that highly active or sedentary adults are more susceptible to cervical disc injury (\cref{fig:case1}(3)). Additionally, the contribution of each modality to the final diagnosis was well demonstrated: the text modality contributed the most, followed by the indicator modality, while the image modality predicted the greatest probability of \textit{``bulging''}, albeit only slightly more than \textit{``'herniated'} (\cref{fig:case1}(6)). The text and image modalities complemented each other, and the prediction of the indicator modality was not critical.

\par \textbf{Takeaway.} E3 and E4 acknowledged that the outcomes align with their clinical background, showcasing the system's ability to carry out inter- and intra-modality analysis. Overall, the physicians concluded that the system adequately fulfilled their demands and resolved their apprehensions, as they remarked, ``\textit{it holds the promise to facilitate the learning process of novice physicians}''.

\subsection{Case II: Learning From Demonstration}
\par Our study involved working with \textbf{I1, I2}, who have two years of academic experience but limited clinical exposure. The main objective was to utilize our system for their educational development by learning from clinical records. During the regulatory training phase, they were mentored by \textbf{E3 -- E4} and received instruction on the fundamental aspects of diagnosing cervical degeneration in an academic setting. Before introducing them to our system, both individuals expressed a keen interest in clinical practice, prompting us to approach this case study with a problem-solving perspective.

\par
\begin{figure}[h]
\centering
\includegraphics[width=\columnwidth]{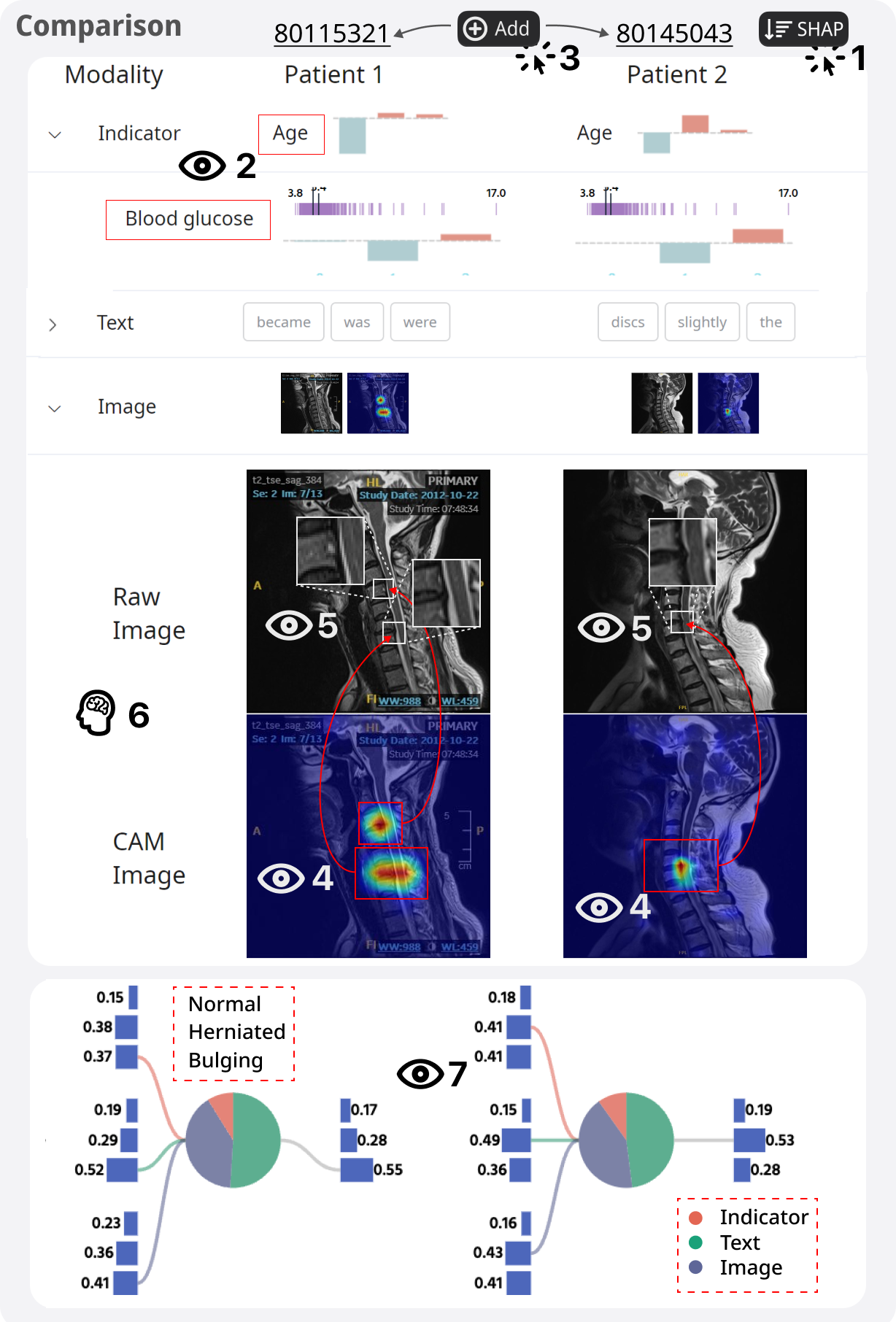}
\vspace{-6mm}
\caption{Case II: Learning From Demonstration. The interns explored the \textit{Comparison View}  to analyze individual patient cases with CardID “80115321” and “80145043”.}
\label{fig:case2}
\vspace{-6mm}
\end{figure}

 During the course of our case study, intern \textbf{I2} posed the question, \textbf{``what are the precise clinical differences between \textit{bulging} and \textit{herniated}?''} She specifically sought to understand these differences in the context of MRI scans, ``\textit{I know bulging to herniated is a transition from mild to severe, but I want to know the exact clinical differences, especially on MRI}". Following our introduction of the dataset, the interns proceeded to explore the fusion embedding. \textbf{I1} then inquired about how to identify the entries for each cluster, which prompted us to remind him of the lasso operation. The interns then proceeded to lasso multiple groups of cases in each cluster, paying particular attention to those with \textit{bulging} and \textit{herniated} entries. They then utilized the \textit{modality exploration view} to examine interpretability-aided records. By selecting the ``Sort by SHAP'' button (\cref{fig:case2}(1)), \textbf{I2} discovered that ``\textit{Age}'' and ``\textit{Blood glucose}'' consistently appeared in the top three indicators (\cref{fig:case2}(2)). She hypothesized that this finding could be attributed to the fact that younger adults are at a higher risk for cervical degeneration, and obesity may be a predisposing factor as indicated by blood glucose levels, ``\textit{this is unexpected, but makes sense}''.

\par The interns further examined the \textit{bulging} and \textit{herniated} groups by selecting the cases with CardID ``80115321'' and ``80145043'' and added these cases to the \textit{Comparison View} (\cref{fig:case2}(3)). They utilized the comparison table for each modality and examined the indicator modality, noting that all indicators were normal. They then shifted to the text modality and observed the highly weighted words ``became'' in the first case and ``slightly'' in the second case. Upon toggling to the image comparison table to scrutinize Raw$\&$CAM images of both cases, the interns utilized the saliency map to identify potential lesion locations. ``\textit{The comparison tables help me a lot, especially the saliency map}'', said \textbf{I1}, ``\textit{it's quite surprising to see the saliency map highlights possible lesion locations, allowing me to know where to check at first glance}''. They noted that the highlighted portion of the saliency map (\cref{fig:case2}(4)) corresponded to the area where the discs (horizontal dark stripes in the Raw image) protruded backward in both cases (\cref{fig:case2}(5)).

\par \textbf{I2} expressed her curiosity about the differences in radiographic detail and commented that ``\textit{both cases had discs protrusion, which I would expect, but what makes the two protrusions different diagnoses?}'' (\cref{fig:case2}(6)). To get an answer, they meticulously compared the two Raw images. They ultimately discovered that the protrusion in the \textit{bulging} case did not extend beyond the vertical white line on the right side, whereas it did in the \textit{herniated} case. The interns then returned to the \textit{modality exploration view} to confirm their understanding, ultimately concluding that the diagnostic basis likely lies in whether the protrusion breaks through the tissue on the right. In the \textit{Probability Fusion View} (\cref{fig:case2}(7)), \textbf{I2} noted that ``\textit{both cases are dominated by images and text, as shown in the pie chart.}''

\par \textbf{Verification of conclusions and clarification of doubts.} After the interns' exploration, we invited both \textbf{E3} and \textbf{E4}, as well as \textbf{I1} and \textbf{I2}, to confirm their findings and address any lingering questions. \textbf{E3} confirmed \textbf{I1}'s conclusion that MRI and radiology reports play a crucial role in our diagnoses, while indicators are often only briefly considered. ``\textit{We do put more emphasis on MRI and radiology reports in our diagnoses. For indicators, we usually just take a glance}''. Additionally, \textbf{E3} confirmed that obese individuals are more likely to experience cervical spine injury due to the excess force on their cervical spine. \textbf{E4} praised the intern's observation, noting that bulging refers to the disc protruding outward while the outer layer of the annulus remains intact, whereas herniated refers to the nucleus spurting out through a tear in the annulus. \textbf{E4} agreed that this aligns with their clinical radiographic judgment. In response to the doubt from \textbf{I2}, \textbf{E4} explained, that ``\textit{in addition to bulging, the upper highlighted area of the MRI indicates ossification of the posterior longitudinal ligament, which is typically not found in cases of cervical degeneration}''.

\subsection{Expert Interview}
\par We conducted a semi-structured interview with all experts for a duration of one hour to obtain their feedback regarding our approach.

\par \textbf{System Performance.} All experts acknowledged the effectiveness of the \textit{DiagnosisAssistant} in enhancing the medical diagnosis learning experience. \textbf{I1} and \textbf{I2} mentioned that before the system's implementation, learning was limited to observing mentors. However, with our system, they can independently explore multimodal data. \textbf{E3} and \textbf{E4} expressed satisfaction with the interpretability techniques used, especially praising the CAM technique for effectively capturing regions of interest in their diagnosis. They also expressed a desire to use these explanations for future investigations. Moving forward, they plan to validate the system through extended real-world usage and supplement it with comprehensive qualitative user studies.

\par \textbf{Visual Designs.} In general, all experts confirmed that the visual representations were intuitive and the system was easy to use. \textbf{E3} remarked that the interface provided \textit{``precisely the information required by a medical professional for performing targeted diagnostic analyses''}. \textbf{I2} noted that \textit{``I am surprised by the interface's functionality, which allowed for multiple modalities of patient data to be presented simultaneously and interconnected''}, stating that it was more intuitive than their usual statistical analysis methods. \textbf{I1} and \textbf{I2} noted that the \textit{Comparison View} was particularly useful for recording their observations.

\par \textbf{Suggestions.} During the exploration, \textbf{I2} and \textbf{E3} detected discrepancies in the data quality, such as inaccuracies in clinical notes. They proposed that the system should automatically propose solutions to rectify such issues. Moreover, the current methods of maintaining data quality are carried out behind the scenes, and physicians desire to be more involved in this process. \textbf{E3} expressed that interactive tools for anomaly detection might be required to supervise the data quality, stating that ``\textit{to ensure data quality, interactive tools for detecting and encoding any missing or incorrect information may be indispensable}''.

\par \textbf{Feedback from External Experts.} With the assistance of our collaborating physicians, we established communication and solicited the proficiency of three external experts specialized in the domain of orthopedics, each possessing a professional tenure surpassing 8 years. Their primary task encompassed the thorough assessment of our system's comprehensive pipeline alongside the corresponding case studies. Consequently, they offered two insightful feedback points for our consideration. First and foremost, the process of clinical diagnosis entails a high level of complexity and multifaceted nature, necessitating meticulous consideration of various symptoms, physical examinations, and test results. In line with recommendations put forth by external experts, the incorporation of a comprehensive range of relevant factors and clinical outcomes within an auxiliary diagnostic system would serve to augment both accuracy and efficiency. Second, given that physicians across different departments often concentrate on distinct facets of a particular disease, it becomes imperative for interns to acquire the skill of integrating their knowledge and experiences derived from diverse backgrounds and disciplines. This integration is essential for effectively managing the intricate nature of clinical practice. Consequently, the integration of AI advancements into an auxiliary diagnostic system, aimed at offering more refined and interdisciplinary perspectives, is likely to enhance the overall comprehension of diseases and foster expertise from multiple perspectives, particularly among interns.

\section{Discussion and Limitation}

\noindent\textbf{Contributions and Effectiveness.} Our approach enhances the learning experience of novice and intern physicians in diagnostic skills by utilizing ML models trained on historical medical data. This approach supports the analysis of medical data across different modalities through interpretability techniques and visual cues, facilitating the acquisition of diagnostic skills. Additionally, our system enables the comparative analysis of individual patients and facilitates advanced exploration of patient characteristics in multimodal scenarios. This approach offers two key benefits: 1) senior clinicians can save time by reducing their deep involvement in training, and 2) novice and intern physicians can learn from objective, data-driven examples from the past, reducing the impact of mentor bias. It is important to note that our approach does not aim to replace the crucial role of senior clinicians in novice and intern training but provides a cost-effective alternative to SBME tools.

\noindent\textbf{Reliability of Post-hoc Explainability Techniques.} Our system utilizes various post-hoc interpretability methods for multimodal models to help users understand model predictions and facilitate learning. However, it is important to note that the explanations provided by the model may not always be accurate, which could lead to incorrect decision-making. Hence, we caution against directly relying on the system for high-stakes clinical decisions. Evaluating the reliability of post-hoc interpretability techniques in clinical settings is beyond the scope of this study and will be explored in future research. Moreover, in conjunction with advancements in human-centered algorithm design and the utilization of more nuanced decision metrics~\cite{sivaraman2023ignore}, we can enhance the acceptability and adoption of AI tools among clinicians.

\noindent\textbf{Generalizability and Scalability.} Although the system possesses the capability to accommodate additional clinical datasets, such as MIMIC-III~\cite{johnson2016mimic}, the cervical spine dataset serves as the principal experimental domain for its evaluation and testing. Our future plans involve incorporating additional types of data, including temporal sequences of patients' vital signs and their genomic profiles, which will necessitate the development of new visualizations and interpretability techniques. Presently, the system is capable of presenting $626$ patient records, including demographic data, test results, clinical notes, and radiological images, effectively fulfilling the needs of physicians. However, the scalability of the system is subject to limitations due to the substantial computational resources demanded by data processing and interpretability algorithms, as well as the current implementation of scatterplots within the \textit{Embedding Transition View}. Furthermore, the display capabilities for images and treemap visualizations may pose constraints on the effective representation of a large number of elements. In the future, we aim to improve computational efficiency by implementing parallel computations. We also plan to enhance the performance of scatterplots by using \textit{deck.gl} on canvas, enabling real-time rendering and smooth interaction with larger patient datasets of up to 10,000 patients. Furthermore, we will investigate the use of hierarchical clustering and semantic zooming as potential solutions to address scalability challenges.

\noindent\textbf{Limitations.} This study has limitations in terms of data quality control, particularly for textual data, and a qualitative user study with a limited number of experts. Managing and analyzing clinical notes poses significant challenges in the field of medical AI. In the future, we intend to conduct further studies to examine the impact of data quality issues on the system's usage. Additionally, we plan to test and enhance the system based on long-term real-world usage. Future evaluations may explore the usability of our system across diverse medical departments and assess its efficacy in continuously updated deployed systems. Additionally, considering the improvement of communication between novice clinicians and our system, there arises a potential need for personalization in human-AI interaction~\cite{calisto2023assertiveness}.

\section{Conclusion and Future Work}
\par In this study, we introduce \textit{DiagnosisAssistant}, a visual analytics system aimed at enhancing the learning experience of interns and novice physicians using historical medical records as a substitute. Through observations and analysis of interactions between experienced physicians and interns/novices, we seek to better understand the ``mentor-apprentice'' process. The system incorporates inter- and intra-modality analysis to visualize multimodal data and integrates a multimodal model into the user interface. We evaluate the system's effectiveness through two case studies and expert feedback. Additionally, we propose a further exploration of reliable post-hoc interpretability techniques for interpreting AI decisions in multimodal clinical scenarios.

\acknowledgments{This work is partially supported by the National Natural Science Foundation of China Youth Fund under Grant No.: 82001471, the Shanghai Frontiers Science Center of Human-centered Artificial Intelligence (ShangHAI), and Key Laboratory of Intelligent Perception and Human-Machine Collaboration (ShanghaiTech University), Ministry of Education.}

\balance
\bibliographystyle{abbrv-doi-hyperref}

\bibliography{template}

\end{document}